\documentclass[twocolumn,amssymb, amsmath, aps, showpacs, footinbib, prb]{revtex4-1}

\usepackage{graphicx}
\usepackage{dcolumn}
\usepackage{bm}
\usepackage{color}

\begin{document}


\title{Thermoelectric properties of two-dimensional slabs of Ba$_{8}$Ga$_{16}$Ge$_{30}$ from 
first principles}

\author{Deepa Kasinathan,$^1$ Vicente Pacheco-Espejel,$^2$ and Helge Rosner,$^1$ }

\affiliation{$^1$Max-Planck-Institut f\"ur Chemische Physik fester Stoffe, 01187 Dresden, Germany}
\affiliation{$^2$ Fraunhofer Institut f\"ur Fertigungstechnik und Angewandte Material Forschung, Institutsteil Dresden, Germany}

\begin{abstract}
Thermoelectric effects enable the direct conversion between thermal and electrical energy and provide an alternative
route for power generation and refrigeration. The clathrate Ba$_{8}$Ga$_{16}$Ge$_{30}$ has the highest
figure of merit ( $ZT$ $\approx$ 1) among other members in the family of type-I inorganic clathrates. 
Enhancement of the thermoelectric properties have been observed in multilayered superlattices, 
quantum wires and in nanostructured materials, either due to
the increase in power-factor ($S^{2}\sigma$) or due to the reduction of lattice thermal conductivity ($\kappa$). 
Here, we investigate the thermoelectric properties of two-dimensional slabs with varying thickness 
of Ba$_{8}$Ga$_{16}$Ge$_{30}$
using semi-classical Boltzmann transport theory with constant scattering approximation. We observe that, there 
exists a delicate balance between the electrical conductivity and the electronic part 
of the thermal conductivity in reduced dimensions and the insights from these results can directly be used to control particle 
size in nanostructuring experiments. The calculated properties are consistent with the recent, first measurements on
bulk nanostructured samples. 

\end{abstract}


\maketitle

\section{Introduction}

Thermoelectric power generators use the temperature gradient to drive the charge carriers (electrons and holes) 
from one end of the material to the other, thereby creating a potential difference. This provides the means of 
converting the temperature gradient directly into electricity without the need for any moving parts. However, 
the inherent low efficiency of the prevailing materials is inadequate to compete with the conventional power 
generation and refrigeration. Hence the central issue in thermoelectrics research is to increase the thermoelectric 
figure of merit $ZT$, a dimensionless quantity defined as $ZT$ = $(S^{2}\sigma/\kappa)/T$, where $S$ is the Seebeck coefficient 
(also known as thermopower), $\sigma$ is the carrier conductivity, $\kappa$ is the total thermal conductivity 
(= $\kappa^{el}$ (electron) + $\kappa^{ph}$ (phonon)), and $T$ is
the absolute temperature. 
As these transport properties depend on interrelated material characteristics, finding thermoelectric materials with 
enhanced $ZT$ values remains a challenging issue.
To enable a more widespread usage of thermoelectric technology power generation and 
 heating/cooling applications, $ZT$ of at least 2-3 is required. Two different research approaches have been proposed 
 for developing the next generation of thermoelectric materials: (i) investigating new families of advanced bulk 
 thermoelectric materials, and (ii) studying low-dimensional or nanostructured material systems.
 The increase in $ZT$ of nanostructured materials could either  be from the reduction of $\kappa^{ph}$ 
 or from the enhancement of the power factor, $S^{2}\sigma$. Hence, a microscopic picture of the transport 
 properties and its response to nanostructuring is necessary to fine tune the preparation techniques. 

\begin{figure}[t]
\begin{center}
\includegraphics[angle=-0,width= 9.0cm,clip]{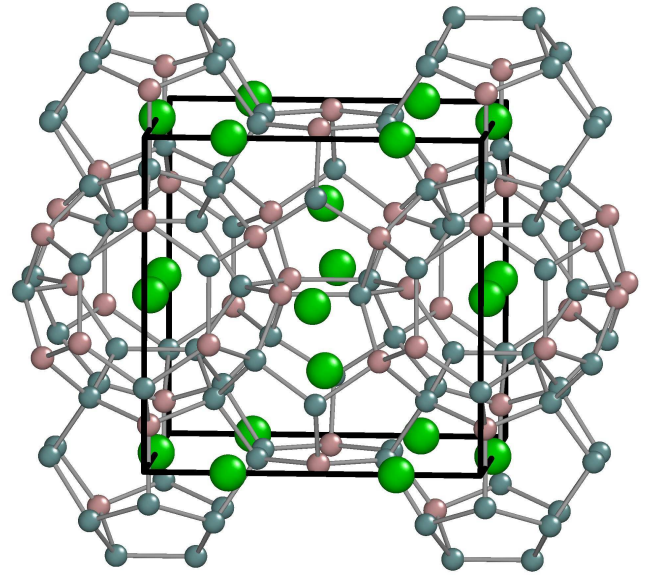}
\caption{\label{str}(Color online) The crystal structure of type-I clathrate Ba$_{8}$Ga$_{16}$Ge$_{30}$.
The large (green) atoms at the center of the cages are the Ba atoms. The framework is made up of two kind 
of cages: a 20-atom smaller pentagonal dodecahedron and a 24-atom larger tetrakaidecahedron. 
The brown atoms are Ga and the grey atoms are Ge.
 }
\end{center}
\end{figure}

Type-I inorganic clathrates are host-guest structures with the guest atoms trapped in the framework of the 
host structure. From a thermoelectric point of view, the clathrates are interesting because they are 
semiconductors with adjustable bandgaps. Investigations in the past decade have shown that type-I clathrates 
$A_{8}$Ga$_{16}$Ge$_{30}$ ($A$ = Ba, Sr, Eu) may have the unusual property of "phonon glass-electron crystal" proposed by 
Slack for good thermoelectric materials.\cite{slack1,slack2} Among the known inorganic clathrates, Ba$_{8}$Ga$_{16}$Ge$_{30}$ has the highest
 figure of merit ($ZT$ $\approx$ 1). Recently, first successful attempts have been made to synthesize nanostructures
 of Ba$_{8}$Ga$_{16}$Ge$_{30}$ using the bottom-up approach.\cite{vicente} The nanoparticles contained mainly thin plates
 that were indexed according to the [100] direction of the clathrate structure type. 
 The measured Seebeck coefficient reaches -145 $\mu$V/K at 648 K, though the $ZT$ values are quite low (0.02) due to 
 the low sample density. Motivated by these measurements, in our work, we concentrate on obtaining a microscopic
 picture of the electronic contributions to the thermoelectric properties for reduced dimensions. 
 In particular, we have considered two-dimensional slabs of varying thickness with the [100] surface termination and
 calculated the thermoelectric properties using the semi-classical Boltzmann transport theory.  
 We observe that, there exists a delicate balance between the electrical conductivity and the electronic part 
 of the thermal conductivity in reduced dimensions and insights from these results can directly be used to control particle 
 size in nanostructuring experiments.

\section{Structure}

The type-I Ba$_{8}$Ga$_{16}$Ge$_{30}$ clathrate has a cubic crystal structure with
the space group $Pm\bar{3}n$. The crystal structure contains 46 tetrahedrally coordinated
host atoms (Ga/Ge) which make two small dodecahedron cages and six larger
tetrakaidecahedral cages (Fig.\,\ref{str}). The heavy guest atoms (Ba) were placed at the center of the 
cages (wyckoff positions 2$a$ and 6$d$).  The Ga atoms were distributed in the framework with 
twelve Ga atoms at 24$k$, one Ga at 16$i$ and three Ga at 6$c$ sites, resulting in only
seven short Ga-Ga bonds with a bond length of $\approx$ 2.48\,\AA. 
Such a distribution of the Ga atoms was done to accommodate the Ga-Ga bond avoidance
observed experimentally.  The remainder of the framework positions were then filled with 
Ge atoms.
The symmetry setting for the calculations was thus lowered to space group 1 to accommodate
such a random distribution of Ga/Ge atoms.

\section{Calculational details}
Non-spin-polarized 
DFT total energy and Kohn-Sham band-structure calculations were 
performed applying the full-potential local-orbital code (version FPLO9.01.35),
within the local 
density approximation (LDA).\cite{fplo1,fplo2}
The Perdew and Wang flavor\cite{PW92} of the
exchange correlation potential was chosen for the scalar relativistic calculations.
The calculations were carefully converged with respect to the number
of $k$ points. 
The  transport properties were calculated using the semiclassical Boltzmann transport 
theory\cite{semi1, semi2, semi3} within the constant scattering approximation as implemented in BoltzTraP.\cite{boltztrap}
This  approximation is based on the assumption that the scattering 
time $\tau$ determining the electrical conductivity does not vary strongly with energy on the 
scale of $kT$. Additionally, no further assumptions are made on the dependence of $\tau$ due
to strong doping and temperature. 
This method has been successfully applied to many narrow band gap materials 
including clathrates and as well as to oxides.\cite{semi3,johnsen2006,kasinathan2007,xiang2007}
One main concern when calculating transport coefficients, is the underestimation
of band gaps using the standard DFT functionals. Such an underestimation of
band gaps manifests in the reduction of thermopower at higher temperatures due
to bipolar conduction. To overcome this problem, we concentrate on analyzing the trend in 
the calculated thermopower for various concentrations of electron and
hole doping rather than quantifying them. 

\begin{figure*}[t]
\includegraphics[width=18cm]{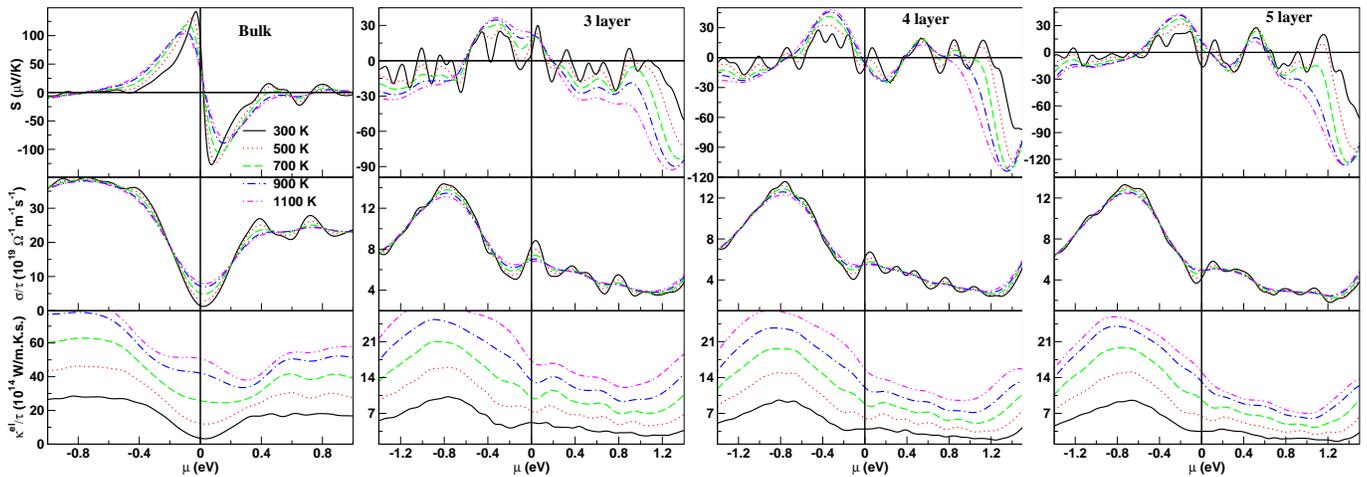}
\caption{\label{transport_compare}
(Color online) Temperature dependence of the calculated thermoelectric properties $S$, $\sigma/\tau$, and $\kappa^{el}/\tau$ as a function of the chemical potential for bulk Ba$_{8}$Ga$_{16}$Ge$_{30}$ along with [100] terminated slabs of varying thickness.
 }
\end{figure*}

\section{Results}

As mentioned previously, the nanoparticles of Ba$_{8}$Ga$_{16}$Ge$_{30}$ synthesized recently contained
mainly nano-plates oriented along [100] direction. Therefore, we begin by considering a three-layer slab
with a [100] surface termination and calculate the thermoelectric properties. Due to the complex nature of the
crystal structure of the clathrates, there is no obvious cleavage plane to choose from when creating the slab 
geometries. Moreover, the cages are all interconnected and every choice of termination will destruct the 
cage network. Hence, we calculated the thermoelectric properties of many three-layer slabs with different 
[100] surfaces. Comparing the values of Seebeck coefficient for the various calculations, we have identified the
termination which results in increasing thermopower as a function of temperature. 
Having identified an advantageous slab geometry, we proceeded to increase the thickness of the slabs and 
evaluated the thermoelectric properties without changing any other structural parameters. 
Collected in Fig.\,\ref{transport_compare} are the Seebeck coefficient, electrical conductivity and the electronic 
part of the thermal conductivity as a function of the chemical potential for a three, four and five layer slab for various
temperatures. The thermoelectric properties of the bulk system is also plotted alongside for easy comparison. 
The bulk system is a Zintl phase and hence a semiconductor. This results in purely electron (chemical potential $\mu$ $>$ 0, $S$ is negative)
or hole (chemical potential $\mu$ $<$ 0, $S$ is positive) conductivity for low carrier concentrations. 
On the contrary, the slab geometries no longer retain the Zintl composition at the surface and hence show 
multiband behavior, witnessed by the change in sign of $S$ for a given $\mu$. 

The values of the calculated thermopower are not affected by
the constant scattering time approximation used to calculate the various
transport properties, since the expression for $S$ is independent of 
$\tau$. This means that $S$ is directly dependent on the electronic 
structure of the material. 
For the three-layer slab, at $\mu$ = 1.2 eV, the charge carriers are mainly electrons and $S$ is increasing with temperature and 
reaches a value of -90 $\mu$V/K at 1100 K.  
Upon increasing the thickness of the slab to four layers, the trend in $S$ is similar and at $\mu$ = 1.2 eV, it reaches
-110 $\mu$V/K at 1100 K. 
Adding an additional layer retains the large values of thermopower. 
Comparing this to the bulk, we can conclude that $S$ remains unaffected (even slightly enhanced) 
upon lowering the dimensions for Ba$_{8}$Ga$_{16}$Ge$_{30}$. 
On the contrary, the electrical conductivity of the slabs reduced by a factor of two for the slab systems in comparison to the
bulk.
Interestingly, the electronic part of the thermal conductivity $\kappa^{el}$ is also reduced with respect to the bulk  for the slab geometries. This term goes into the denominator of the expression that defines $ZT$, and a lower 
$\kappa^{el}$ for the slabs means an increase in $ZT$ values with respect to the bulk. 
Assuming that nanostructuring reduces the phononic part of the thermal
conductivity $\kappa^{ph}$ by introducing grain boundaries, one can obtain improved 
thermoelectric performance in nanostructured Ba$_{8}$Ga$_{16}$Ge$_{16}$ clathrates.

\section{Conclusions}
We have calculated the electronic contribution to the thermoelectric properties for two-dimensional slabs of 
Ba$_{8}$Ga$_{16}$Ge$_{30}$ clathrate from semi-classical Boltzmann statistics. We discern a dependence of the
different surface terminations and slab thickness on the calculated thermoelectric properties. 
A delicate balance exists between $S$, $\sigma$ and $\kappa^{el}$ in reduced dimensions. 
Enhanced values of $S$ is obtained for heavily $n$-doped slabs of Ba$_{8}$Ga$_{16}$Ge$_{30}$ with respect to
the bulk. Both $\sigma$ and $\kappa^{el}$ are reduced for the slabs in comparison to the bulk, but 
$S^{2}\sigma/\kappa^{el}$ of the slabs are comparable to the best bulk values. 
Assuming a reduction of $\kappa^{ph}$ via nanostructuring, one can anticipate improved 
values of $ZT$ for low dimensional Ba$_{8}$Ga$_{16}$Ge$_{30}$ systems. \\

Acknowledgement: This work was supported by the Deutsche Forschungsgemeinschaft DFG in the frame of a priority program SPP1386 Ò$Nanostrukturierte Thermoelektrika: Theorie, Modellsystemme und kontrollierte Synthese$Ó with the grant number PA 1821/1-1.

\end{document}